\documentclass[acmsmall]{acmart}

\usepackage[ruled]{algorithm2e}
\usepackage{graphicx}
\usepackage{textcomp}
\usepackage{xcolor}
\usepackage{listings}
\usepackage{makecell}
\usepackage{threeparttable}
\usepackage{subfigure}
\usepackage{url}
\usepackage{mdframed}
\usepackage{multicol}
\usepackage{tcolorbox}
\usepackage{enumitem}
\usepackage{booktabs}
\usepackage{multirow}
\usepackage{pifont}
\usepackage{amsmath}
\usepackage{subfigure}

\definecolor{forestgreen}{rgb}{0.25,0.63,0.21}

\lstdefinelanguage{JavaScript}{
  keywords={typeof, new, true, false, catch, function, return, null, catch, switch, var, if, in, while, do, else, case, break, export, static, const, let, class, void, async, await},
  basicstyle=\ttfamily\small\bfseries\footnotesize,
  keywordstyle=\bfseries,
  ndkeywords={class, export, boolean, throw, implements, import, this},
  ndkeywordstyle=\color{darkgray}\bfseries,
  identifierstyle=,
  sensitive=false,
  frame=none,
  numbers=left,
  xleftmargin=2em,
  comment=[l]{//},
  morecomment=[s]{/*}{*/},
  commentstyle=\color{gray}\ttfamily,
  stringstyle=\color{purple}\ttfamily,
  morestring=[b]',
  morestring=[b]"
}

\lstdefinelanguage{diff}{
    basicstyle=\ttfamily\bfseries\footnotesize,
    comment=[s]{/*}{*/},
    keywordstyle=\bfseries,
    morecomment=[f][\color{red!75}]{-},
    morecomment=[f][\color{forestgreen!75}]{+}
}

\newtheoremstyle{customstyle1}{}{}{}{}{\bfseries}{.}{ }{\thmname{#1}\thmnumber{ #2}\thmnote{: #3}}

\newtheoremstyle{customstyle2}{}{}{}{}{\scshape}{.}{ }{\thmname{#1}\thmnumber{ #2}\thmnote{: #3}}

\theoremstyle{customstyle1}
\newtheorem{reason}{Reason}

\theoremstyle{customstyle2}

\AtBeginDocument{%
  \providecommand\BibTeX{{%
    \normalfont B\kern-0.5em{\scshape i\kern-0.25em b}\kern-0.8em\TeX}}}

\setcopyright{acmlicensed}
\copyrightyear{2018}
\acmYear{2018}
\acmDOI{XXXXXXX.XXXXXXX}

\acmJournal{JACM}
\acmVolume{37}
\acmNumber{4}
\acmArticle{111}
\acmMonth{8}

\begin{document}

\title{Towards Better Comprehension of Breaking Changes in the NPM Ecosystem}


\author{Dezhen Kong}
\authornote{Both authors contributed equally to the paper.}
\affiliation{%
  \institution{The State Key Laboratory of Blockchain and Data Security, Zhejiang University}
  \city{Hangzhou}
  \country{China}}
\email{timkong@zju.edu.cn}

\author{Jiakun Liu}
\authornotemark[1]
\affiliation{%
  \institution{School of Information Systems, Singapore Management University}
  \city{Singapore}
  \country{Singapore}
}
\email{jkliu@smu.edu.sg}

\author{Lingfeng Bao}
\authornote{Corresponding author.}
\affiliation{%
 \institution{The State Key Laboratory of Blockchain and Data Security, Zhejiang University}
 \city{Hangzhou}
 \country{China}}
\email{lingfengbao@zju.edu.cn}
\additionalaffiliation{
    \institution{Hangzhou High-Tech Zone (Binjiang) Blockchain and Data Security Research Institute}
}

\author{David Lo}
\affiliation{%
  \institution{School of Information Systems, Singapore Management University}
  \city{Singapore}
  \country{Singapore}}
\email{davidlo@smu.edu.sg}

\renewcommand{\shortauthors}{Dezhen Kong, et al.}

\begin{abstract}
Code evolution is prevalent in software ecosystems, which can provide many benefits, such as new features, bug fixes, security patches, etc., while still introducing breaking changes that make downstream projects fail to work. Breaking changes cause a lot of effort to both downstream and upstream developers: downstream developers need to adapt to breaking changes and upstream developers are responsible for identifying and documenting them. In the NPM ecosystem, characterized by frequent code changes and a high tolerance for making breaking changes, the effort is larger. 

For better comprehension of breaking changes in the NPM ecosystem and to enhance breaking change detection tools, we conduct a large-scale empirical study to investigate breaking changes in the NPM ecosystem. We construct a dataset of explicitly documented breaking changes from 381 popular NPM projects. 
We find that 95.4\% of the detected breaking changes can be covered by developers' documentation, and 19\% of the breaking changes cannot be detected by regression testing.
Then in the process of investigating source code of our collected breaking changes, we yield a taxonomy of JavaScript and TypeScript-specific syntactic breaking changes and a taxonomy of major types of behavioral breaking changes.
Additionally, we investigate the reasons why developers make breaking changes in NPM and find three major reasons, i.e., to reduce code redundancy, to improve identifier names, and to improve API design, and each category contains several sub-items.

We provide actionable implications for future research, e.g., automatic naming and renaming techniques should be applied in JavaScript projects to improve identifier names, future research can try to detect more types of behavioral breaking changes. By presenting the implications, we also discuss the weakness of automatic renaming and breaking change detection approaches, such as the lack of support for public identifiers and various types of breaking changes.

\end{abstract}

\ccsdesc[300]{Software and its engineering~Software libraries and repositories}
\ccsdesc[500]{Software and its engineering~Software evolution}

\keywords{Breaking Change, NPM, JavaScript, Code Evolution}

\received{20 February 2007}
\received[revised]{12 March 2009}
\received[accepted]{5 June 2009}

\maketitle

\section{Introduction}\label{section:introduction}

The evolution of code is prevalent in software ecosystems~\cite{konstantopoulos2009best, raemaekers2012measuring, lehman1980programs, traini2021software}. Developers of upstream software libraries make code changes to incorporate new features, bug fixes, security patches, component refactorings, and extra-functional improvements \cite{traini2021software, tao2012software}. However, code evolution may break the contract previously established with its downstream by introducing \textit{breaking changes} (BCs) in its public APIs, making client applications fail to work \cite{xavier2017historical}. For example, renaming frequently used methods or classes can make client projects that depend on the code fail to compile. Therefore, downstream developers will be required to make efforts to adapt to such breaking changes. 

Considering the risks and the effort brought by breaking changes to downstream developers, many works have been utilized to help developers analyze and measure the impacts of breaking changes \cite{mezzetti2018type, moller2019model,Dont_Break}.
However, they relied on test suites in downstream projects, and either detected JavaScript breaking changes by directly running test suites \cite{Dont_Break}, or via API models\footnote{Intuitively, an API model defines the type restrictions of an API \cite{mezzetti2018type}.} generated from dynamic execution of test cases \cite{mezzetti2018type, moller2019model}.
This leads to the result that breaking changes observed in prior studies are located in well-known APIs that are popularly used in downstream projects, such as \texttt{Client.socket} in \textsf{socket.io}\footnote{\url{https://github.com/socketio/socket.io/commit/b73d9be}}, \texttt{Request.prototype.onResponse} in request\footnote{\url{https://github.com/request/request/commit/d05b6ba}}, and \texttt{each} in async.
{However, breaking changes are reported to upstream projects because they already manifested themselves in downstream test cases or already caused bugs.}
For example, in version 3.0.0 of a famous JavaScript library \textsf{lodash}, the behavior of function \texttt{mixin} was changed and not explicitly marked by upstream developers, and a downstream developer reported this issue by self-constructed test code.\footnote{\url{https://github.com/lodash/lodash/issues/880}}
Hence, completely relying on test case execution is not sufficient for understanding breaking changes.

We observe that in the NPM ecosystem, a number of projects write commit messages complying with Conventional Commits \cite{Conventional_Commits}.
According to Conventional Commits, {BREAKING CHANGE} tokens are documented in commit messages to indicate that the corresponding commits contain breaking changes. This is because the core developers of upstream libraries need to ensure the stability of the API and protect the reputation of the upstream projects. When a new code change is submitted for review, they carefully inspect it for the presence of breaking changes, as well as other bugs or defects. By leveraging the documented breaking changes provided by developers, there is a potential opportunity to notify downstream users about these breaking changes.

However, developers experience increased pressure within the NPM ecosystem, primarily driven by frequent code changes and a high tolerance for breaking changes \cite{bogart2021and,bogart2016break}.
Another possible factor is that different developers may have different perceptions of breaking changes and some of them may ignore the breaking changes, leading to unaware breaking changes that cause unforeseen bugs in downstream projects. If we could characterize the large number of documented breaking changes at source code level, it would assist upstream developers in gaining a deeper understanding of the types of commits that could potentially introduce breaking changes. Furthermore, such categorization would provide valuable insights for researchers to develop tools aimed at detecting breaking changes in the future.

To bridge this gap, we conduct an empirical study to better comprehend documented breaking changes. We first select popular projects with over 50 GitHub stars and associated links to an existing GitHub repository, resulting in a total of 35,786 repositories. We then randomly select 381 repositories (95\% confidence level and 5\% margin of error) for further investigation. We then clone their associated repositories. After this, we identify BC-related commits admitted by developers by searching for {BREAKING CHANGE} tokens in commit messages. For the 5,242 identified BC commits, we manually select 1,519 commits that 1) do not contain too long commit messages, 2) are related to JavaScript production code and 3) are associated with documentation that can explain the reasons behind the commit. These selected BC commits are distributed across 131 distinct projects and the projects can be grouped into six categories according to functionalities and usages, i.e., utilities, frontend projects, development-related tools, database-related tools, plugins and Web API related tools (detailed in Section \ref{section:methodology}), which makes the BC-related commits in our projects representative. To extract breaking changes from these BC commits, we build breaking change type taxonomy by learning from a previous API detection tool for Java \cite{brito2018apidiff} and adding new JavaScript and TypeScript-specific features. Our taxonomy includes \textsc{Remove}, \textsc{Rename}, \textsc{Change Signature}, \textsc{Change Behavior} types, etc., which is detailed in Section \ref{section:breaking_change_types}. Through several thematic analyses, we obtain JavaScript and TypeScript specific code features (RQ2 and RQ3) and a taxonomy of reasons behind breaking changes. We answer the following research questions in this study:

\textbf{RQ1: {To what extent do \textit{detected breaking changes} and \textit{documented breaking changes} overlap?}} In our work, we collect breaking changes from developers' intention, i.e., the explicitly documented breaking changes. We first check whether documented breaking changes actually break test code. Therefore, we do regression testing on our collected BC-related commits, similar to previous works \cite{mezzetti2018type, moller2019model, mostafa2017experience, zhang2022has}.
We find 95.4\% of detected breaking changes are documented and 81\% of the documented breaking changes can be detected by regression testing. The result shows that most detected breaking changes are well documented, while a proportion of documented breaking changes cannot be detected by regression testing. Since the documented breaking changes cover most of the detected breaking changes, it is reasonable to extract breaking changes from documentation (especially commit messages, issues, and pull requests).

\textbf{RQ2: {What syntactic breaking changes in the NPM ecosystem are specific to JavaScript and TypeScript?}}
We investigate whether there are some breaking changes related to JavaScript-specific language features. Despite the triviality of most syntactic breaking changes (such as moving classes to another package, removing a field of a class), we find some JavaScript-specific BCs are notable, including two removing operations, i.e., 1) removing default export, 2) removing export of an element in one place, and five types of JavaScript and TypeScript specific signature changes, e.g., parameter changes in configuration objects, switch between callback and Promise, etc.

\textbf{RQ3: How do developers make behavioral breaking changes?} In this RQ, we investigate how developers perform breaking changes at source code level. Since syntactic breaking changes can be detected through syntax analysis and code refactoring detection, we mainly focus on behavioral breaking changes (a.k.a, semantic breaking changes in some works) that change the internal program logic. We find four major types of \textsc{Change Behavior} breaking changes, i.e., 1) changing the specification return values, 2) changing process for some option values, 3) changing default or initial value of variables and 4) changing error handling method.

\textbf{RQ4: {Why do developers make breaking changes in NPM ecosystem?}}
In this RQ, we revisit the rationale of breaking changes in the NPM ecosystem and find three main factors (each with some sub-factors) that motivate developers to make breaking changes, i.e., 1) to reduce code redundancy, 2) to improve identifier names, 3) to improve API design, and divide each reason into some sub-items, which extends the previous works \cite{bogart2016break, bogart2021and, brito2020you}.

Based on our empirical findings, we provide actionable implications for future research, including 1) automatic renaming deserves much concern in JavaScript projects since poor identifier naming is a main contributor to technical debt, 2) automatic tools for ensuring code consistency are necessary, 3) future research should strive to detect more types of behavioral breaking changes. By presenting these implications, we also discuss the weakness of automatic renaming and breaking change detection approaches, such as lacking support for public identifiers and various types of breaking changes.

The contributions of our paper are two-fold:

\begin{enumerate}
    \item We build a carefully constructed breaking change dataset extracted from a wide range of JavaScript and TypeScript projects in the NPM ecosystem.
    \item We empirically identify how developers perform breaking changes at source code level, yielding notable findings like JavaScript and TypeScript specific syntactic breaking changes and typical actions of behavioral breaking changes. We also investigate why developers perform breaking changes in the NPM ecosystem.
\end{enumerate}

The remainder of this paper is organized as follows. Section \ref{section:background} introduces preliminary knowledge and motivating examples. Section \ref{section:methodology} presents the methodology of our study. Section \ref{section:results} details our empirical findings. Section \ref{section:discussion} provides our implications for future research from this study. Section \ref{section:threats_to_validity} discusses the threats to validity. Section \ref{section:related_work} reviews the related work and Section \ref{section:conclusion} concludes our work. 

\noindent \textbf{Data availability.} We provide the replication package of our research at \url{https://doi.org/10.5281/zenodo.13927690}.

\section{Background}\label{section:background}

In this section, we describe some preliminary knowledge on breaking changes.

\subsection{Preliminary of Breaking Changes}

Prior to our work, a number of studies have uncovered breaking changes from many aspects, such as the occurrence of breaking changes, the impacts of breaking changes on downstream projects, and motivations for making breaking changes.
The research works \cite{decan2019package, venturini2023depended, raemaekers2017semantic} have demonstrated the widespread occurrence of breaking changes in the NPM ecosystem and their effects on client applications. Venturini et al. found that in their sampled packages, 11.7\% of all client packages and 13.9\% of their releases are impacted by breaking changes, and notably, 44\% of the breaking changes are introduced in minor or patch releases (according to Semantic Versioning \cite{Semantic_Versioning}, developers should not introduce breaking changes in non-major releases) \cite{venturini2023depended}.
Bogart et al. found that coarse-grained motivations for making breaking changes include requirements and context changes, bugs and new features, rippling effects from upstream changes, and technical debt from postponed changes \cite{bogart2016break, bogart2021and}. 

However, despite much knowledge of BC, to the best of our knowledge, none of the studies precisely defined BC. Researchers collected breaking changes in various forms in previous studies. For example, in the evaluation of several breaking change detection tools \cite{zhang2022has, mezzetti2018type}, the authors collected breaking changes by running downstream test cases: if a test case of a downstream project could not pass with a newer provider (after the code change), then the code change was identified as a BC. However, on the one hand, not all providers are dependent on many client applications. If a code change has no triggering test case in client applications, we cannot determine whether it is actually incompatible. On the other hand, client code may not follow the specification of providers and incorrectly access APIs, e.g., passing an improper value to a function, which will result in undefined behavior \cite{moller2019model}, including test failure, but it does not reflect a BC. In another breaking change detection tool {APIDiff} \cite{brito2018apidiff}, Brito et al. utilized the code refactoring detection tool {RefDiff} \cite{silva2017refdiff} to identify syntax-related breaking changes (e.g., changing method signature or renaming class). However, as they pointed out, the detected breaking changes are just {breaking change candidates} (BCC), since some syntax changes are applied to internal methods and classes, which are not intended for public use. To this end, they also asked developers to check whether the detected BCCs are actually breaking changes \cite{brito2020you}.

In our study, we adopt Brito et al.'s criteria \cite{brito2020you}, since we also investigate the breaking changes from developers' perspective, i.e., how and why they perform breaking changes. Specifically, a breaking change should 1) be confirmed by developers, 2) be categorized into pre-defined BC types.
To achieve the first criterion, we identify breaking changes by searching explicit BC declarations in the documentation (usually commit messages). To achieve the second criterion, we define several breaking change types for JavaScript and TypeScript on the basis of Brito et al.'s Java BC types \cite{brito2018apidiff} and JavaScript development experience. We detail the BC types in Section \ref{section:breaking_change_types} and describe how to extract breaking changes in Section \ref{section:bc_extraction_analysis}.

\subsection{Types of Breaking Changes in NPM Projects}\label{section:breaking_change_types}

\begin{table}[htbp]
    \centering
    \renewcommand{\arraystretch}{1.15}
    \caption{Breaking Change Types Adopted from Previous Works}
    \label{table:supported_refactoring_actions}
    \resizebox{0.9\linewidth}{!}{
    \begin{threeparttable}
    
    \begin{tabular}{ll}
    \toprule
       {Type} & {Supported Elements} \\
       \midrule
        \textsc{Rename} & module, class, interface, enum, type, method, field, constant, variable \\
        \textsc{Remove} & module, class, interface, enum, type, method, field, constant, variable  \\
        \textsc{Move} & module, class, interface, enum, type, method, field, constant, variable  \\
        \textsc{Inline} & method  \\
        \textsc{Push Down} & method, field  \\
        \textsc{Change Signature} & method, field  \\
        \textsc{Change Behavior} & the content of method, field, constant and variable  \\
        \bottomrule
    \end{tabular}
    \end{threeparttable}
    }
\end{table}

Existing BC detection tools for Java, typically {APIDiff}~\cite{brito2018apidiff}, support many syntactic and object-oriented programming (OOP) related BC types, such as \textit{remove classes} and \textit{push down fields}. Since OOP features are also supported in JavaScript (since ECMAScript 2015 \cite{ECMAScript_2015}, \texttt{class} can be directly used while legacy code must leverage the prototype mechanism \cite{inheritance_and_prototype_chain}), we learn from the breaking change types for the Java language and adapt them to JavaScript and TypeScript (a superset of JavaScript, in practice many source code files are written in TypeScript and will be compiled into JavaScript files) by retaining available BC actions in JavaScript and adding more breaking change types which are not considered previously. We then construct a taxonomy of JavaScript and TypeScript breaking changes (shown in Section \ref{table:supported_refactoring_actions}), where \textit{interface}, \textit{enum} and \textit{type} are only available in TypeScript. We explain each type as follows:

\begin{enumerate}[leftmargin=2em]
\item \textbf{\textsc{Rename}} refers to identifier changes of public code elements.  \textsc{Rename} BCs can be applied to many code elements, such as exported classes, interfaces, and enums, as well as methods and fields in exported classes and interfaces.

\item \textbf{\textsc{Remove}} refers to removal of public code elements. \textsc{Remove} BC can be applied to many code elements. After a \textsc{Remove} BC, the affected public element is not accessible by downstream applications, e.g., an exported class is completely removed, or no longer marked as exported.

\item \textbf{\textsc{Move}} can be applied to all APIs supported by \textsc{Rename} BCs. For example, a class is moved to another package since its functionality is more related to that package.

\item \textbf{\textsc{Inline}} is a refactoring operation in object-oriented programming: removing a public method and copying its body into an existing method. Simple methods are often inlined to make code more straightforward. In fact, Inline can be regarded as a special case of \textsc{Remove}, since the inlined public method is removed from public access.  By contrast, normal “Remove” actions do not copy the content of the removed method to another existing method. However, since inline is a very common refactoring operation in object-oriented languages, we still regard it as a separate breaking change type, like previous works \cite{brito2018apidiff, silva2017refdiff}.

\item \textbf{\textsc{Push Down}} is also a refactoring operation in object-oriented programming: moving the method in the parent class into the child class since this method is only used in one child class in reality. Push down can be regarded as a special case of \textsc{Remove} BC.

\item \textbf{\textsc{Change Signature}} refers to the modification of method signatures except for directly changing method names. Adding or removing modifiers (such as ``{static}'', ``{private}'') and changing the parameters are typical refactoring actions of this type.

\item \textbf{\textsc{Change Behavior}} (also called \textit{semantic breaking changes} in some works \cite{mezzetti2018type, zhang2022has}) refers to those operations changing the internal behavior (e.g., programmatic logic in methods, and content of global constants that can be used by some methods) rather than syntactic elements, such as changing class names and adding required parameters. In other words, after a behavioral breaking change is performed, the affected classes (or methods, interfaces, etc.) can be invoked as the same way as before.

\end{enumerate}

In our work, we focus on breaking changes \emph{in JavaScript source code}, hence we do not take breaking changes in non-source code files into consideration, typically CSS, Markdown, HTML, \texttt{package.json} files and configuration files related to TypeScript, Webpack and ESLint, etc.

\section{Methodology}\label{section:methodology}

\subsection{Data Collection}\label{section:data_collection}

Due to the vast number of JavaScript projects in NPM, our study focuses on selecting the most popular projects to ensure that the collected breaking changes are representative.
\textcolor{black}{To do so, we utilize the Libraries.io open source repository and dependency metadata provided by Reid et al. \cite{reid_2020_3898749} to retrieve the most popular JavaScript and TypeScript projects since Libraries.io is commonly used in prior studies \cite{he2021large, decan2019empirical, decan2018impact}. We first select popular projects having more than 50 GitHub stars and associated with an existent GitHub repository, which yields 35,786 repositories. We then randomly select 381 repositories (95\% confidence level and 5\% margin of error) for further investigation.}

\begin{figure}
    \centering
    \includegraphics[width=0.9\linewidth]{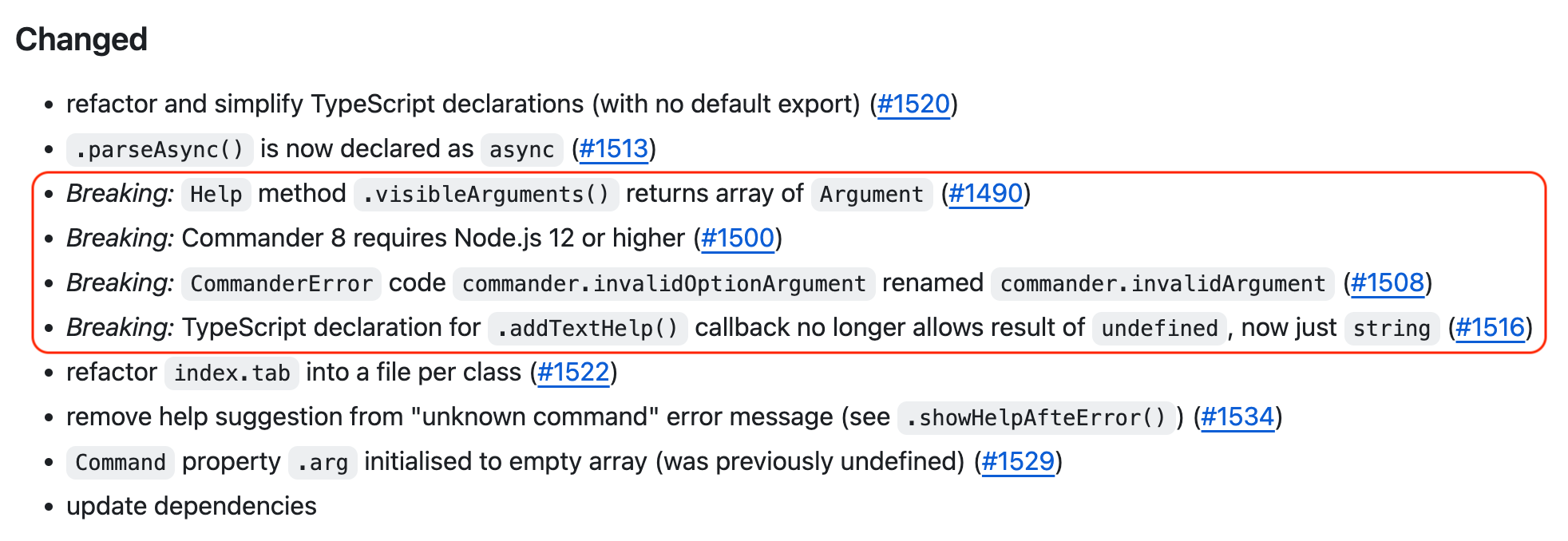}
    \caption{Screenshot of the Changelog of commander.js 8.0.0 (The four highlighted breaking changes can be linked to related issues and commits)}
    \label{figure:1}
\end{figure}

Since we consider the breaking changes that are explicitly confirmed by developers, we try to extract breaking changes from developers' documentation, typically commit messages, \textit{changelogs}, issues and pull request text on GitHub.
We notice that Conventional Commits \cite{Conventional_Commits} provides a standard for writing readable commit messages. According to the standards, if a commit message contains a \texttt{BREAKING CHANGE} section, the commit is identified as a breaking change. For example, the text below shows a typical commit message indicating a breaking change:

\begin{lstlisting}[frame=none]
refactor: compiler -> runtimeCompiler
BREAKING CHANGE: compiler option has been renamed to runtimeCompiler
\end{lstlisting}

We first check whether Conventional Commits are widely used, analyzing the commits from the 381 projects we used. We find that 360 out of 381 repositories contain commits that conform to Conventional Commits and 198 projects have over 80\% commits that follow Conventional Commits. We use the regular expression below to check Conventional Commit compliance:

\begin{lstlisting}
(fix|feat|chore|build|ci|test|style|perf|refactor)(\([a-zA-Z0-9-\s]+\))?!?:
\end{lstlisting}

Here \texttt{fix}, \texttt{feat}, \texttt{chore}, etc., are the defined or recommended scope tokens in Conventional Commits specification. This indicates that \textit{Conventional Commits} specification is widely used and we can use it as a simple way to extract plenty of breaking changes. Considering that many projects do not follow this standard, we also try our best to extract BC commits from \textit{changelogs} of each project (if developers have declared), issues and pull requests on GitHub. A \textit{changelog} summarizes the changes from the last version release, and may mention multiple breaking changes. By analyzing the \textit{changelog}, we can link the mentioned breaking changes to corresponding commits. For example, the project \textsf{tj/commander.js} does not use Conventional Commits, then we consult the mentioned pull request IDs in \texttt{CHANGELOG.md} file\footnote{\url{https://github.com/tj/commander.js/blob/master/CHANGELOG.md}} (shown in Figure \ref{figure:1}, where four breaking changes are mentioned), and we also analyze the content of the pull requests to obtain the commits related to the breaking changes. In this way, we obtain 5,242 commits.

Then two of the authors manually inspect them and remove some of them that are not related to JavaScript production code (e.g., only modifying package.json, Markdown documentation, build scripts, test cases) or contain very long commit messages (over 10 lines) that are difficult to understand. For example, if a commit only states that ``{drop support for Node.js 14}'' and modifies the corresponding item in \texttt{package.json}, it will be discarded. Additionally, we do not include the commits that contain multiple breaking change declarations. Specifically, we remove 1,826 commits that contain over 10 line commit messages or multiple BREAKING CHANGE declarations and 692 commits that are not JavaScript code changes. Finally, we retain 2,724 candidate commits.

\subsection{Regression Testing}\label{section:regression_testing}

We use regression testing to detect breaking changes in our selected projects (381 in total) to check:

\begin{enumerate}
    \item what percentage of detected breaking changes are well documented,
    \item what percentage of documented breaking changes can be detected by regression testing.
\end{enumerate}

To check \ding{172}, due to the number of the total commits being too large (1,005,344), we first sample 16,371 commits for further analysis (99\% confidence level and 1\% margin of error). We then remove the commits in these projects that only contain non-JavaScript code changes (e.g., test code changes, Markdown and HTML documentation changes, dependency updates in \texttt{package.json}). Then for each commit $c$ and its prior commit $c^{\prime}$ we do regression testing following the steps below:

\begin{enumerate}
    \item We run the test cases in $c^{\prime}$ directly (using \texttt{npm test} command). We discard the commits that cannot be configured (e.g., there are version conflicts during installation).
    \item We restore the production code from $c$. The production code refers to JavaScript and TypeScript source code files not located in \texttt{test} directory.
    \item We run the test cases (using \texttt{npm test} command).
    \item If the first step succeeds and the third step fails, then we say that $c$ is a detected BC commit.
    \item If a detected BC commit is documented with BREAKING CHANGE, or can be found in \textit{changelog}s, issues, or pull requests, we regard the commit as a documented BC commit.
\end{enumerate}

We use the test cases written by developers rather than dependents since 1) we have found that 363 out of the 381 projects provide test suites, and developers' test cases can better express the intended usage of APIs \cite{dinella2022toga}, while many of the 381 projects do not have plenty of dependents, 2) some projects are not intended for pragmatic use, especially CLI tools in our projects such as \textsf{npm/cli}, therefore no clients access the APIs in these projects. To check \ding{173}, we also apply the steps described above to the 2,724 BC-related commits from Section \ref{section:data_collection}. 

\subsection{Breaking Change Selection and Analysis}\label{section:bc_extraction_analysis}

\subsubsection{Breaking Change Selection and Categorization} We select feasible breaking changes for further research from the 2,724 candidate BC commits and categorize them into types described in Section \ref{section:breaking_change_types}. Since we need to understand \textit{what} a commit changed and \textit{why} developers made it, we require the commit message and other documentation of a commit (including issues, pull requests and \textit{changelogs}) to contain some explanation of \textit{why developers made that commit}. We here reuse Tian et al.'s taxonomy of \textit{why} information in commit messages \cite{tian2022makes}. Typical \textit{why} information includes:

\begin{enumerate}
\item Issue description: about the linked issue, weakness of current code implications, etc.
\item Requirements Illustration: about the usage need, out-of-date statements, etc.
\item Objective description: objectives, such as fixing bugs, improving performance, refactoring code, etc.
\item Necessity Implication: relation to prior commits, benefits of making such code changes, etc.
\end{enumerate}

For example, considering the commit \underline{cebd670a} of \textsf{angular/angular}\footnote{\url{https://github.com/angular/angular/commit/cebd670a}}, the developers just stated renaming the method \texttt{requestCheck} in \texttt{ChangeDetectorRef}  to \texttt{markForCheck}. By inspecting the related issue (\#3403), the developer said that ``{When I first saw the requestCheck() method on ChangeDetectorRef, I assumed it was how I manually run change detection. Others have assumed this as well}'', which is a description of ``Necessity Implication''.

If there is no \textit{why} information covering the aspects mentioned above, we discard the commit. Two of the authors independently check whether the documentation contains reason-related information (Cohen’s Kappa value is 0.80) and determine the type of each commit (using the taxonomy in Section \ref{section:breaking_change_types}, Cohen's Kappa value is 0.77). Then the authors hold meetings to solve the disagreements. In total we identify 1,519 commits that contain reason-related information.

\subsubsection{Breaking Change Distribution}

\begin{figure}
    \centering
    \includegraphics[width=0.78\linewidth]{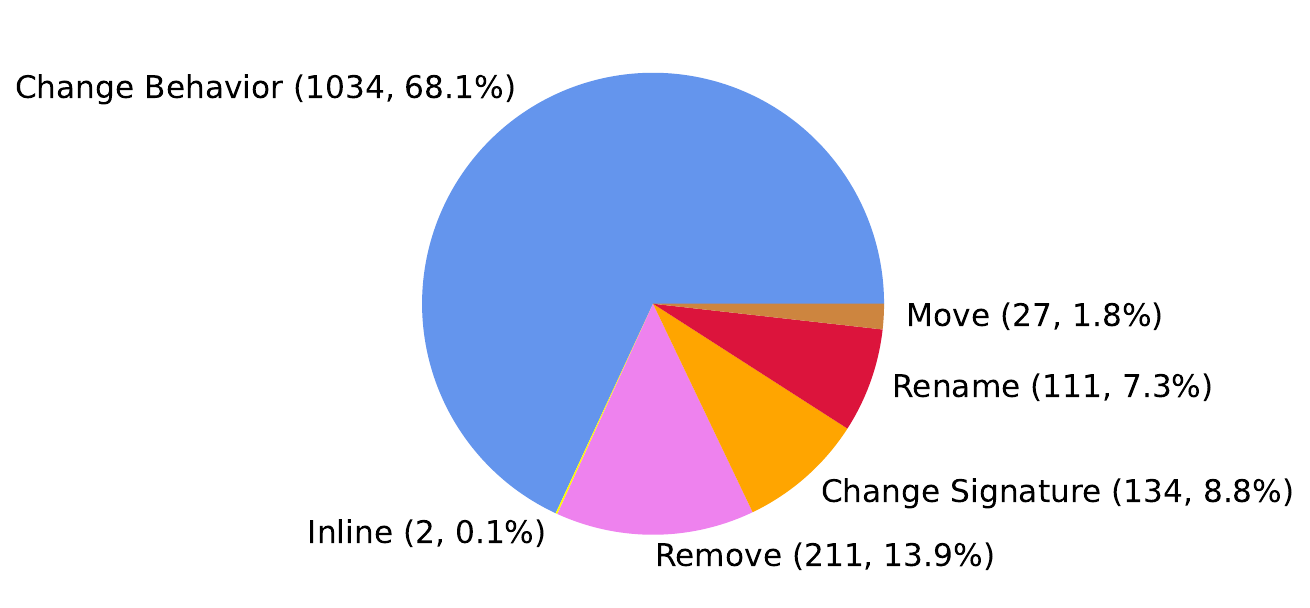}
    \caption{Categories of Our Investigated Breaking Changes}
    \label{figure:2}
\end{figure}

Figure \ref{figure:2} presents the percentages of different types of breaking changes.
It can be obviously seen that \textsc{Change Behavior} BCs make up the largest proportion of all BC types (68.1\%). By contrast, \textsc{Move} and \textsc{Inline} BCs make up the smallest percentage (1.8\% and 0.1\% respectively). In our further investigation, we mainly focus on \textsc{Change Behavior} BCs.

\begin{table}[htbp]
    \centering
    \renewcommand{\arraystretch}{1.15}
    \caption{NPM Projects Containing Breaking changes Used in Our Study}
    \label{table:projects}
    \resizebox{0.75\linewidth}{!}{
    \begin{tabular}{llr}
    \toprule
       {Category} & {Description} & {Number} \\
    \midrule
       Frontend  & Used in Web browser environment  & 45 \\
       
       Web API & Providing access to RESTful APIs & 9 \\
       Database Tools & Providing database operations & 4 \\
       Development Tools & Providing project management & 25 \\
       Plugin & Served as plugins for other NPM projects & 3\\
       Utility & Other useful JavaScript libraries & 45 \\
       \bottomrule
    \end{tabular}
    }
    
\end{table}

\begin{table}[htbp]
    \centering
    \caption{Proportion of Involved Projects of Each Breaking Change Category}
    \label{table:project_count_bc}
    \resizebox{0.95\linewidth}{!}{
    \begin{tabular}{lrrrrrrr}
    \toprule
       BC Category  & Frontend & Utility & Development & Database & Web API & Plugin & \# All \\
    \midrule
      \textsc{Remove}  &  10/45 & 12/45 & 8/25 & 4/4 & 4/9 & 0/3 & 38/131 \\
      \textsc{Rename} & 10/45 & 10/45 & 3/25 & 3/4 & 3/9 & 0/3 & 29/131\\
      \textsc{Move} & 2/45 & 2/45 & 1/25 & 1/4 & 1/9 & 0/3 & 7/131 \\
      \textsc{Inline} & 0/45 & 0/45 & 1/45 & 0/4 & 0/9 & 0/3 & 1/131 \\
      \textsc{Change Signature} & 8/45 & 10/45 & 6/25 & 3/4 & 2/9 & 0/3 & 29/131 \\
      \textsc{Change Behavior} & 44/45 & 41/45 & 25/25 & 4/4 & 9/9 & 3/3 & 130/131 \\
      \bottomrule
    \end{tabular}
    }
\end{table}

The projects that contain breaking changes in our study serve various types of functionalities, including command-line interface utilities, REST API SDKs and Web frontend frameworks. They can be grouped into six categories (shown in Table \ref{table:projects}). For example, \textsf{pnpm/pnpm}\footnote{\url{https://github.com/pnpm/pnpm}} is a alternative CLI tool for official NPM implementation, \textsf{renovatebot/renovate} provides automatic project build and \textsf{aws/aws-cdk}\footnote{\url{https://github.com/aws/aws-cdk}} wraps the AWS cloud APIs, \textsf{gajus/eslint-plugin-flowtype}\footnote{\url{https://github.com/gajus/eslint-plugin-flowtype}} is an ESLint plugin, and async/async\footnote{\url{https://github.com/async/async}} is a fundamental utility of asynchronous programming. Table \ref{table:project_count_bc} shows the number of involved projects for each type of breaking changes.

\subsubsection{Breaking Change Labeling}

For the 1,519 remaining commits, two of the authors conducted two thematic analyses to analyze \ding{172} source code features (for RQ2 and RQ3) and \ding{173} reasons behind the commit (for RQ4). The guideline recommended by Cruzes et al. \cite{cruzes2011recommended} were used. The guideline for the two analyses is shown below. Two of the authors independently performed the steps above, and they held a series of meetings to solve the agreements.

\begin{enumerate}
\item The authors read the related documentation, especially commit messages carefully to understand what the developers want to express.

\item The authors read the documentation again and generate phrases as initial codes that describe the source code features, how developers make breaking changes and the reasons behind the commit.

\item The authors aggregated the codes with similar meanings and generate a theme name to describe each cluster.

\item The authors then reviewed all themes and try to merge the themes with similar semantics, or made the similar themes become sub-items of a new theme.
\end{enumerate}

After obtaining all themes, the authors assigned possible themes to each breaking change. Note that for simplicity, we assigned each retained breaking change with one reason, which is consistent with the previous works on motivations behind breaking changes \cite{brito2020you}. In this process, the authors held several meetings to resolve the disagreements on the reason behind each commit since the reasons behind some BC commits are difficult to determine. For example, in the commit \underline{2713380} of \textsf{reactivex/rxjs}\footnote{\url{https://github.com/reactivex/rxjs/commit/2713380}}, developers renamed \texttt{inspect} to \texttt{audit}, the reasons behind this rename operation was presented in Issue \#1505\footnote{\url{https://github.com/reactivex/rxjs/issues/1505}} and \#1387\footnote{\url{https://github.com/reactivex/rxjs/issues/1387}}. Developers mentioned two possible reasons, i.e., name collision (since \texttt{inspect} conflicts with \texttt{util.inspect} provided by Node.js\footnote{See explanation in Issue 1387 of Rxjs.}) and synonym replacement (since \texttt{audit} is synonymous to \texttt{inspect} and it is less confusing). By carefully reading and understanding the related issues, two of the authors reach an agreement and assign the reason ``to avoid name conflict'' to this breaking change, because if \texttt{inspect} were not used in Node.js native objects, the identifier \texttt{inspect} is still acceptable and does not need to be renamed. In the process of assigning the reason to the breaking changes, the authors have very few disagreements (the Cohen's Kappa value 0.94).

\section{Results}\label{section:results}
In this section, we present our empirical results for each research question.

\subsection{RQ1: To What Extent do Detected Breaking Changes and Documented Breaking Changes Overlap?}\label{section:results_rq1}

From 16,371 commits used in regression testing in Section \ref{section:regression_testing}, we detect 173 BC commits (1.0\%), 95.4\% of them (165) are documented by developers. And for 2,724 BC commits, 2,206 (81\%) can be detected by regression testing. The results illustrate that most detected breaking changes are well documented, while a proportion (19\%) of documented breaking changes cannot be detected via regression testing since developers might forget to write test cases or current test cases are too general to cover the code modifications.

The undocumented breaking change commits are possibly due to \ding{172} the commit is made too long ago when Conventional Commits was not proposed, and \ding{173} the project does not comply to Conventional Commits. And the main reasons that regression testing cannot detect breaking changes is that the BC behavior can only be triggered by external conditions (e.g., network error), hence it is difficult to simulate such a situation. For example, in commit \underline{61e7a81a} of \textsf{octokit/rest.js}, the \texttt{followRedirects} option is no longer supported. This BC can only manifest itself when the HTTP server returns a status code between 301 and 307. However, this is not often seen, and developers have not written any test case to cover the process of ``followRedirects'' option. Developers might forget to write test cases and current test cases might be too general to cover the code modifications, hence automatic test case generation may be useful.

For the 19\% of breaking changes that cannot be detected, since developers in popular projects have rich development experience, we believe those breaking changes are trustable and worth attention. Although some breaking changes currently do not affect many downstream projects, they should still be pointed out since in some cases downstream developers may write code that invokes the broken API, then the downstream developers can confirm the breaking changes quickly by looking up commit messages. For example, in commit \underline{669592d} of \textsf{socketio/socket.io}\footnote{\url{https://github.com/socketio/socket.io/commit/669592d}} contains a breaking change that removed \texttt{Socket\#binary} method. It is not detected when we run regression testing. However, after several months, one developer asked for an alternative method of \texttt{Socket\#binary}\footnote{https://github.com/socketio/socket.io/discussions/3826}. This indicates that although an API in a breaking change is not frequently used, and currently no test cases can trigger it, the API is  possibly invoked at a certain time.

\subsection{RQ2: What Syntactic Breaking Changes in NPM Ecosystem Are Specific to JavaScript and TypeScript?}\label{section:results_rq2}

While most code-level actions in breaking changes are also available in other programming languages, such as changing parameter orders and removing classes, etc., we highlight the noticeable syntactic BC actions in our investigated projects.

\subsubsection{JavaScript-specific remove operations} In ECMAScript 6, a module can have many exported items and a default export item. Therefore, developers can not only remove the code of a class, interface or enum, etc. like other OOP languages, but also just remove them from exported list and put them into internal source code files. We present the JavaScript-specific \textsc{Remove} operations in our collected breaking changes as follows:

\noindentparagraph{\textbf{Remove a default export (6 cases).}} ECMAScript 6 supports default export functionality\footnote{\url{https://developer.mozilla.org/en-US/docs/Web/JavaScript/Reference/Statements/export}}. For example, in commit \underline{102e4b0} of \textsf{nodkz/mongodb-memory-server}, developers simply removed default export of the module \textsf{util}. Therefore, the code \texttt{import generateDbName from \textquotesingle{util}\textquotesingle} will not work.

\noindentparagraph{\textbf{Remove an export position (17 cases).}} For example, in commit \underline{cfbfaac} of \textsf{reactivex/rxjs}, the types in \texttt{rxjs/interfaces} modules are no longer accessible and users must use them by importing them from the main module. Therefore, users can import \texttt{UnaryFunction} with the following two methods, while after the code change, they can only import \texttt{UnaryFunction} using the second form.

\begin{lstlisting}[language=JavaScript]
import { UnaryFunction } from 'rxjs/interfaces';
import { UnaryFunction } from 'rxjs';
\end{lstlisting}

\subsubsection{JavaScript-specific signature changes} Most signature changes in JavaScript are also available in other languages like Java, e.g., adding, removing, and reordering parameters. However, some JavaScript-specific signature changes should be focused on. We present the categories as follows:

\noindentparagraph{\textbf{About parameters in configuration objects (15 cases).}} In JavaScript, many properties can be packaged in one configuration object. Considering the \textit{diff} shown below, the second parameter of the method \texttt{userinfo} is an object, and this function only extracts four properties in this object using destructuring assignment syntax \cite{destructuring_assignment}. After the code change, the \texttt{verb} property is renamed to \texttt{method}, since \texttt{GET} is an HTTP method. Correspondingly, the internal code related to this property is also changed.

\begin{lstlisting}[language=diff]
   async userinfo(accessToken, 
-    { verb='GET', via='header', tokenType, params } = {}
+    { method='GET', via='header', tokenType, params } = {}
   ) { /* ... */ }
\end{lstlisting}

\noindentparagraph{\textbf{About \texttt{this} parameter (6 cases).}} In the old implementation, the method accepts a callable object (typically a function) and \texttt{this}, and binds the function to \texttt{this}\footnote{\texttt{bind} is a native function in JavaScript. See \url{https://developer.mozilla.org/en-US/docs/Web/JavaScript/Reference/Global_Objects/Function/bind}}. After the code change, the method no longer provide this parameter, the user should first bind the callable object to \texttt{this}. For example, in the code snippet shown below, \texttt{findIndex} only accepted \texttt{predicate} parameter.

\begin{lstlisting}[language=diff]
    export function findIndex<T>(
      predicate: (value: T, index: number, source: Observable<T>) => boolean,
-      thisArg?: any
    ): OperatorFunction<T, number> {
-     return operate(createFind(predicate, thisArg, 'index'));
+     return operate(createFind(predicate, undefined, 'index'));
   }
\end{lstlisting}

\noindentparagraph{\textbf{About \textit{sync} and \textit{async} (20 cases).}} In JavaScript, asynchronous functions can be implemented by using callback arguments and Promise objects. Legacy JavaScript asynchronous functions are designed with callback-style interface, while new implementations often use \texttt{async} directly. For example, in the commit \underline{8c3cecae} of \textsf{automattic/mongoose}, developers provided multiple signatures of an API before the code change, e.g., \texttt{createIndex} have the following function signature:

\begin{lstlisting}[language=JavaScript]
createIndexes(options: mongodb.CreateIndexesOptions, callback: CallbackWithoutResult): void; 
createIndexes(callback: CallbackWithoutResult): void;
createIndexes(options?: mongodb.CreateIndexesOptions): Promise<void>;
\end{lstlisting}

After the code change in this commit, the signature 1 and 2 were removed. The code \textit{diff} below shows this pattern:

\begin{lstlisting}[language=diff]
-  SomeFunction(param, (returnValue, err) => {
-    // Some code about returnValue
-  }
+  let returnValue = await SomeFunction(param);
+  // Some code about returnValue
\end{lstlisting}

Another case is converting synchronous to asynchronous (or vice versa), e.g., simply adding \texttt{async} modifier to the function signature.

\noindentparagraph{\textbf{About {undefined} and {null} (8 cases).}} The keyword {null} represents an intentional empty value, while {undefined} occurs in accessing uninitialized variables or non-existent object properties, and should not be explicitly assigned to an object. The \textit{diff} below shows this pattern:

\begin{lstlisting}[language=diff]
-  public get(key: Key): Value | null {
-    return this.collection.get(key) || null;
+  public get(key: Key): Value | undefined {
+    return this.collection.get(key);
   }
\end{lstlisting}

However, in JavaScript projects without \texttt{d.ts} files, there are also signature related changes, e.g., a method no longer accepts a null value, or a null value might cause error after a code change. In this case, developers cannot represent the change in method signature, therefore, we classify the case as behavioral change.

\noindentparagraph{\textbf{Switch between ``required'' and ``optional'' of fields in method or interface declarations (3 cases).}} TypeScript script projects and part of JavaScript projects use \texttt{d.ts} files to declare interface and method signature. For example, if one parameter is changed to be required, then a ``?'' will be removed from the parameter type. The \textit{diff} below (from commit \underline{86074a6} of coinbase/rest-hooks) shows the signature change type:

\begin{lstlisting}[language=diff]
   static url<T extends typeof SimpleResource>(
     this: T,
-    urlParams?: Partial<AbstractInstanceType<T>>,
+    urlParams: Partial<AbstractInstanceType<T>>,
   ): string {
-   if (urlParams) {
-      if (
-       Object.prototype.hasOwnProperty.call(urlParams, 'url') &&
-       // ...
\end{lstlisting}

\subsection{RQ3: How do Developers Make Behavioral Breaking Change?}\label{section:results_rq3}
We identify four main types of behavioral changes in our collected JavaScript projects of NPM ecosystem. Each type is presented as follows:

\noindentparagraph{\textbf{Changing the specification of return values {(79 cases)}.}} One typical case is that the type of returned objects are changed. For example, in commit \underline{b99f6d3} of \textsf{automattic/mongoose}, the method \texttt{MongooseArray.map()} returned a plain JavaScript array rather than a headless Mongoose array:

\begin{lstlisting}[language=diff]
   map() {
-    const ret = super.map.apply(this, arguments);
-    ret[arraySchemaSymbol] = null;
-    ret[arrayPathSymbol] = null;
-    ret[arrayParentSymbol] = null;
+    const copy = [].concat(this);

-    return ret;
+    return Array.prototype.map.apply(copy, arguments);
   }
\end{lstlisting}

\noindentparagraph{\textbf{Changing the process of some options (231 cases).}} Suppose that one possible value of an option (can be in JSON configuration or parameters) is not supported, then the code that gets properties from the option will be changed, and the corresponding branch of a specific option value will be removed. For example, in commit \underline{ad4f1493} of project \textsf{octokit/rest.js}\footnote{\url{https://github.com/octokit/rest.js/commit/ad4f1493}} (shown below), the judgment of \texttt{options.type} is changed, and \texttt{netrc} branch in a {switch} structure is also removed to completely remove support for \texttt{netrc} mechanism. 

\begin{lstlisting}[language=diff]
   if (!options.type || 
-   'basic|oauth|client|token|integration|netrc'.indexOf(options.type) === -1
+   'basic|oauth|client|token|integration'.indexOf(options.type) === -1
    ) {
       throw new Error("Invalid authentication type, must be" + 
-         'basic', 'integration', 'oauth', 'client' or 'netrc'")
+         'basic', 'integration', 'oauth', or 'client'")
   }
\end{lstlisting}

In CLI tool projects, developers often change the process of some options to remove or change the purpose of an option. For example, in commit \underline{f6fd0c3} of \textsf{pnpm/pnpm}, the option \texttt{-{}-store-path} is not an alias of \texttt{-{}-store} anymore. Therefore, developers remove the code of handling the option, shown as follows:

\begin{lstlisting}[language=diff]
  // in function run (argv: string[]) 
    await new Promise((resolve, reject) => {
      setTimeout(() => {
-       if (opts.storePath && !opts.store) {
-         logger.warn('the store-path config is deprecated.')
-         opts.store = opts.storePath
-       }

      // `pnpm install ""` is going to be just `pnpm install`
      const cliArgs = cliConf.argv.remain.slice(1).filter(Boolean)
      // more code ...
\end{lstlisting}

\noindentparagraph{\textbf{Changing the default behavior for unprovided values (203 cases).}} The arguments of a function can often affect the behavior. When the value of a parameter is not provided (i.e., being \texttt{undefined}), developers may design special program logic to deal with this case, or make the parametre still remain undefined. Sometimes the default program logic of dealing with such case can be changed. As an example, in commit \underline{72bbda7f} of \textsf{npm/cli}\footnote{\url{https://github.com/npm/cli/commit/72bbda7f}}, the default value of local variable \texttt{depthToPrint} is set to zero while it is initially \texttt{undefined}. After this change, the code is as follows:

\begin{lstlisting}[language=JavaScript]
const { /* ... */, depth, /* ... */ } = npm.flatOptions;
const depthToPrint = all ? Infinity : (depth || 0);
// more code
\end{lstlisting}

The default values can be also in global configurable objects. For example, Renovate puts configurations that affect the behavior in \textsf{lib/config/options/index.ts}, and the modification to it will cause behavioral change.

\noindentparagraph{\textbf{Changing error handling process (42 cases).}} One case is changing change the content or format of the error objects. For example, in commit \underline{dd6306e3} of \textsf{octokit/rest.js}\footnote{\url{https://github.com/octokit/rest.js/commit/dd6306e3}}, the error message is parsed as a JSON object and the properties will be put into error object (the existing \texttt{message} property will be overwritten). Since developers usually parse \texttt{error.message}, this change will make the code like \texttt{JSON.parse(error.message)} not work. The code \textit{diff} below shows this change:

\begin{lstlisting}[language=diff]
  // in a Promise chain
  .then(data => { /* ... */ })
  .catch(error => {
     if (error instanceof HttpError) {
+      try {
+        Object.assign(error, JSON.parse(error.message))
+      } catch (_error) {
+        // ignore, see #684
+      }
       throw error
   }
\end{lstlisting}

Another case of this category is changing the criteria of detecting errors, e.g, some status values are no longer regarded as success. In the commit \underline{201f189d} in \textsf{angular/angular}\footnote{\url{https://github.com/angular/angular/commit/201f189d}}, if the return code of an XHR (XML HTTP Request) is not 200, an error will occur, while before the code change, if the return codes between 200 and 300 (exclusive) were all regarded as success states. The breaking change in \texttt{xhr\_backend.ts} is shown as follows:

\begin{lstlisting}[language=diff]
   let response = new Response(responseOptions);
+  if (isSuccess(status)) {
     responseObserver.next(response);
     responseObserver.complete();
+    return;
+  }
+  responseObserver.error(response);
\end{lstlisting}

The code in \texttt{http\_util.ts} that contains the new anonymous function \texttt{isSuccess}:

\begin{lstlisting}[language=diff]
+  export const isSuccess = 
+    (status: number): boolean => (status >= 200 && status < 300);
\end{lstlisting}

We also compare the behavioral breaking changes to those in the Maven ecosystem. Zhang et al. have studied the behavioral breaking change types in Maven projects \cite{zhang2022has}. They found that changing execution logic and changing calculation of the output make up over 60\% of the total breaking changes collected by them. They also studied benign changes that will not cause incompatibilities, such as additional/changed/deleted conditions and branches (over 55\% of the total benign changes), additional try-catch statements, and assignment revision. However, in our findings about source code patterns of JavaScript behavioral breaking changes, it is very common to change conditions in ``if'' statements to remove support for some option values or to change evaluation criteria of some results, although it seems benign. Hence, in JavaScript, the condition changes that involve option/configuration values deserve attention since they are more probable to be breaking changes.

\subsection{RQ4: Why do Developers Make Breaking Changes in NPM Ecosystem?}\label{section:results_rq4}

\begin{figure}[htbp]
    \centering
    \includegraphics[width=0.96\linewidth]{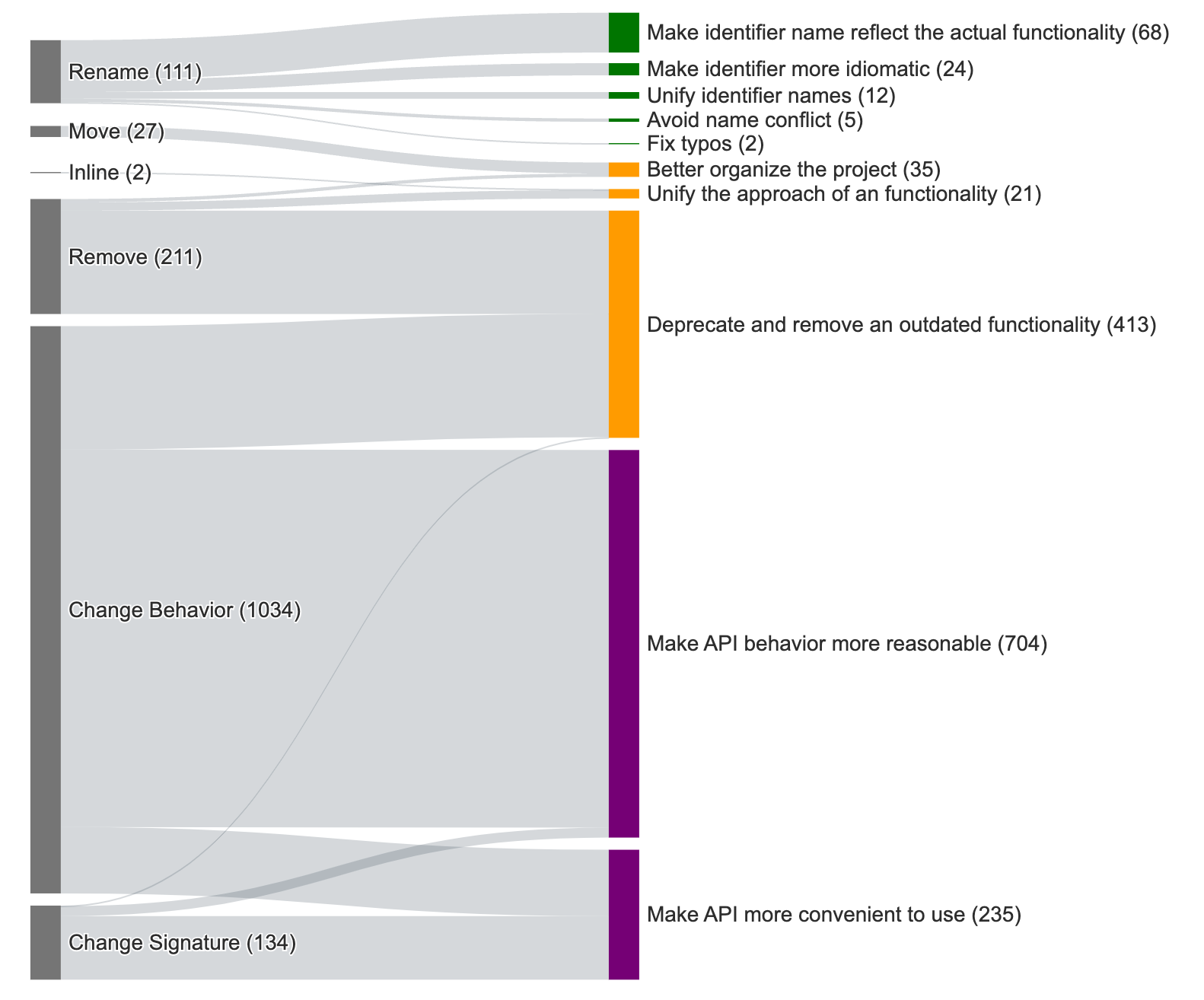}
    \caption{Distribution of Motivations in Different Types of Breaking Changes}
    \label{figure:3}
\end{figure}

In this section, we summarize six major motivations behind making breaking changes by carefully analysis of our collected breaking changes. We show the connection between BC types and motivations in Figure \ref{figure:3}.

\begin{reason}[To reduce code redundancy]\label{reason:1}
(469 cases) Some source code is no longer needed, hence developers remove them. There are two cases of this reason:

\begin{itemize}[leftmargin=1em]
\item To unify the approach of a functionality (21 cases). The functionality can be achieved by other methods, or by using other packages. Therefore, the current implementation can be removed. For example, in commit \underline{1354171} of \textsf{thi-ng/umbrella}, developers removed function \texttt{writeFile}, and recommend using \texttt{writeFile} provided by another package \textsf{@thi-ng/rstream-log-file} since the third-party function can also achieve the same functionality.

\item {To deprecate and remove an outdated functionality (413 cases).} Some outdated input (e.g., authentication mechanism) are deprecated thus developers remove their support. For example, in commit \underline{ad4f1493} of \textsf{octokit/rest.js}, the support for \texttt{netrc} authentication method was removed while others were retained.

\item {To better organize the project (35 cases).} Moving part of the functionality out and create a new package will make the project better organized. For example, in commit \underline{0fa2830}\footnote{\url{https://github.com/jsdoc/jsdoc/commit/0fa2830}} of \textsf{jsdoc/jsdoc}, the package \textsf{@jsdoc/core} was reorganized and some methods in \textsf{@jsdoc/core} moved to a new package \textsf{@jsdoc/cli}.

\end{itemize}
\end{reason}

\begin{table}[]
    \centering
    \renewcommand{\arraystretch}{1.15}
    \caption{Typical \textsc{Rename} Operations}
    \label{table:rename_operations}
    \resizebox{0.96\linewidth}{!}{
    \begin{tabular}{lp{0.7\linewidth}r}
    \toprule
       \textbf{Category}  & \textbf{Explanation} & \textbf{Number} \\
       \midrule
    
        {Add token} & add non-trivial tokens, e.g., from description to ariaDescription & 23 \\
        {Change token order} & change orders in the identifier, e.g.,  secretJsonValue to secretValueFromJson & 2 \\
        {Remove token} & remove one token from the identifier, e.g., renderBoundElementIndex to boundElementIndex & 15 \\
        {Replace} & completely replace the token with another token, e.g., signed to trusted & 20 \\
        {Replace token} & replace one token in identifier, e.g., fromDockerHub to fromDockerRegistry & 39 \\
        {Trivial} & add \texttt{is} or \texttt{get} prefix, e.g., from failed to isFailed & 12 \\
        \bottomrule
    \end{tabular}
    }
    
\end{table}

\begin{reason}[To improve identifier names]\label{reason:2}
(111 cases) Good identifier names can indicate the functionalities. However, improper identifier names can be a main contributor to technical debt in JavaScript projects. Improper identifier names may come from lack of familiarity with external knowledge and a failure to consider previous code. This reason can be divided into the following categories:

\begin{itemize}[leftmargin=*]
    \item {To make the identifier name reflect the actual functionality (68 cases).} The name of a class, interface or method, etc., should reflect the actual functionality. Therefore, after a code change, the identifier names tend to be modified. For example, in issue \#3403 of \textsf{angular/angular}\footnote{\url{https://github.com/angular/angular/issue/3403}}, the developer complaint that ``{When I first saw the requestCheck() method on ChangeDetectorRef, I assumed it was how I manually run change detection. Others have assumed this as well}''. Hence after the issue was reported, the identifier \texttt{requestCheck} is changed to \texttt{markForCheck}.

    \item {To make the identifier more idiomatic (24 cases).} For example, adding prefix \texttt{is} or \texttt{get} to the original name (e.g., from \texttt{empty} to \texttt{isEmpty}), and changing the order of tokens in an identifier (e.g., from \texttt{listSecretsForRepo} to \texttt{listRepoSecrets}).

    \item To unify identifier names (12 cases). Similar classes or methods (e.g., components with the same style in a package) should share a similar name. For example, in commit \underline{9bbd2469} of \textsf{pnpm/pnpm}\footnote{\url{https://github.com/pnpm/pnpm/commit/9bbd2469}}, the field \texttt{localPackages} was renamed to \texttt{workspacePackages}. In the dependency \textsf{@pnpm/resolver-base}\footnote{This package is part of pnpm, and the \textsc{Rename} BC is located in the package @pnpm/find-workspace-packages, also a part of pnpm.}, \texttt{LocalPackages} had been renamed to \texttt{WorkspacePackages}, hence to make it consistent, the field was also renamed.

    \item {To avoid name conflict (5 cases).} For example, in commit \underline{da19583} of \textsf{adonisjs/adonis-framework}\footnote{\url{https://github.com/adonisjs/adonis-framework/commit/da19583}}, developers renamed \texttt{Context} to \texttt{HttpContext} since \texttt{Context} is a commonly used keyword.

    \item {To fix typos (2 cases).} For example, renaming \texttt{MetricAarmProps} to \texttt{MetricAlarmProps}.
\end{itemize}

We also compared the rename operations in JavaScript (and TypeScript) with those in Java since most Java and JavaScript projects adopt the \textit{camelCase} naming convention. For JavaScript and TypeScript projects, we show the rename operations in Table \ref{table:rename_operations}. For Java projects in Maven ecosystem, we inspect Huang et al.'s dataset on Java API migration \cite{huang2021repfinder}. Interestingly, we find that most method rename BCs are trivial, i.e., only adding or removing prefixes \texttt{get}, \texttt{is} (e.g., from \texttt{failed} to \texttt{isFailed}), and only few classes are renamed with one token replaced added, removed or replaced. Other popular languages like Python, Go and Rust, do not use \textit{camelCase} convention.

\end{reason}

\begin{reason}[To improve API design]\label{reason:3}
(939 cases) The quality of API can impact the productivity of programmers, the adoption of APIs, and the quality of dependent code \cite{myers2016improving}. This reason can be divided into the following sub-categories:

\begin{itemize}[leftmargin=1em]

\item {To make API behavior more reasonable (704 cases).} The major scenarios of API's unreasonable behavior include: \ding{172} The API design does not conform to a certain standard and users may misuse the API. For example, in the commit \underline{201f189d} of angular/angular, after the code change, status codes less than 200 and greater than 299 will cause error, while previously, errors only occurred when network errors. \ding{173} The API behavior can easily cause unintended results. For example, in commit \underline{7d4c399} of \textsf{gajus/eslint-plugin-jsdoc}, the behavior is changed unless a new option is set to true. This is to decrease the false positives when capitalized letters on newlines merely represent proper nouns.

\item {To make APIs more convenient to use (235 cases).} For example, changing the parameter order of a method can be more in line with users' habits, and using \texttt{async} rather than \texttt{Promise} or callback parameters also provides convenience for downstream developers.

\end{itemize}

\end{reason}

\section{Implications}\label{section:discussion}

On the basis of the results above, we can yield the following implications for future works:

\noindent \textbf{Automatic naming and renaming techniques can be applied in NPM projects.}  In Section \ref{section:results_rq3}, we conclude that the main reason for renaming identifiers is \textit{to make identifier name reflect the actual functionality}. Therefore, on the one hand, poor identifier naming is a noticeable technical debt in JavaScript projects, hence automatic naming can be utilized to provide good identifier names, and on the other hand, in some cases, renaming identifiers is reasonable due to the necessary adaptation to frequent code changes in NPM ecosystem, thus we suggest applying automatic renaming techniques to make identifier names adapt to code evolution.

Several rule-based, static analysis-based, and machine learning-based approaches are proposed to help automatic naming and renaming. For example, the tool proposed by Caprile et al. \cite{caprile2000restructuring} is a typical rule-based approach that makes unified tokens in an identifier comply to syntax rules, the tool proposed by Feldthaus et al. \cite{feldthaus2013semi} is a static analysis-based approach to semi-automatically rename object properties with a focus on related {property identifiers}\footnote{Property identifiers refer to property names of an object, e.g., if \texttt{option} is a parameter, in \texttt{options.followSymLink}, \texttt{followSymLink} is a property identifier.}, and RefBERT \cite{liu2023refbert} is a typical deep learning-based approach that utilizes fune-tuning pre-training model (i.e., RoBERTa) and infers \textit{local variable} names with contextual token sequences. However, these approaches suffer from two major limitations, and the future works can try to mitigate them:

\begin{itemize}[leftmargin=1em]
\item They are not proposed to deal with identifiers that can be directly accessed by downstream developers (e.g., class and interface names), while \textsc{Rename} actions on these identifiers make up the most in JavaScript projects. For instance, the most recent renaming approach RefBERT works in the same way as code completion, i.e., the intra-function context is input to infer local variable names, however, inferring identifier names needs different context knowledge, such as class structure, client invocation code, etc. 
\item They are not code-change-aware. Since we have found that many a proportion of \textsc{Rename} breaking changes are performed to ensure code consistency (Figure \ref{figure:3} in Section \ref{section:results_rq3}), a renaming technique had better leverage knowledge in prior related code changes to recommend proper names.
\end{itemize}

\noindent \textbf{Tools for detecting code with similar functionalities are necessary.} According to Figure \ref{figure:3}, a number of \textsc{Remove} breaking changes are motivated by approach for unification, i.e., provide a unified API for a specific functionality. However, the relationship between these two or more code changes is indirect and not simple to detect. Besides, in our BC extraction process, we find developers make other small changes for better performance, compatibility, and robustness though they are unrelated to breaking changes. This also to some degree interferes with BC identification.
The findings in our investigation can provide the following directions for future research:

\begin{itemize}[leftmargin=1em]
\item Investigate how to indicate potential BC actions when one BC has been already performed and recommend similar BC actions to improve code consistency and reduce code redundancy.
\item Identify functionally related code snippets that can be removed together to facilitate clean code.
\item Distinguish non-breaking changes from breaking changes and ordinary code changes to provide more convenience for developers.
\end{itemize}

\noindent \textbf{Future research can focus on detecting more types of behavioral breaking changes.} Existing approaches for JavaScript behavioral breaking change detection are NoRegrets \cite{mezzetti2018type} and its enhanced version NoRegrets+ \cite{moller2019model}, to the best of our knowledge. In Section \ref{section:introduction}, we have highlighted their reliance on test cases in downstream projects. In addition, the two approaches can only detect type-related breaking changes. Specifically, in the first phase, they run client test cases and monitor the flows of values by program instrumentation to build API models (intuitively, type restriction of APIs), and in the second phase, they rerun the client test cases with updated upstream code to check whether type restrictions are changed during execution. Therefore, they cannot detect non-type-related breaking changes, e.g., a BC that changes the semantics of the returned string, while the return type is still string. According to Section \ref{section:results_rq3}, there are four types of behavioral BC actions, where only \textit{changing the specifications of return values} can directly cause type changes, while other behavioral changes may not change object types during execution. Therefore, we recommend future works try to deal with various types of behavioral breaking changes uncovered by this study.

\section{Threats to Validity}\label{section:threats_to_validity}

\textbf{Threats to internal validity} are mainly related to the clarity of developers' documentation of breaking changes. Few developers might have not explicitly documented the breaking changes in commit messages, and commit messages (as well as text in issues and pull requests) might not fully reflect developers’ reason for making breaking changes. Also, the discussion and documentation related to breaking changes may not fully reflect developers' consideration. Second, in our manual analysis, we might still not understand developers' intentions since many breaking changes' commit messages do not clearly explain what and why they make breaking changes. Additionally, we might regard some changes in code \textit{diff} as BC-related actions since they are confusing. To address this potential threat, we have double-checked the analysis results to ensure the quality. 

\textbf{Threats to construct validity} are concerned with the errors during breaking commit selection. In our study, we mainly consider the commits that follow Conventional Commits specification and collect breaking change commits that contain BREAKING CHANGE token, which may miss the commits that not fully follow the specification but are actually breaking. To mitigate this, we try our best to collect breaking changes from software documentation, especially \textit{changelogs}, issues and pull requests (mentioned in Section \ref{section:data_collection}). Additionally, we might overlook some commits, although they satisfy our selection criteria although we carefully inspect each collected commits (in Section \ref{section:bc_extraction_analysis}).

\textbf{Threats to external validity} refer to the generalizability of our study. In our study, while we involve 1,519 breaking changes extracted from 381 randomly sampled projects from a vast number of popular NPM projects, and these projects cover diverse application fields of JavaScript language, such as utilities, frontend, Web APIs, database, etc., the extracted breaking changes only account for a small proportion of breaking changes in the whole NPM ecosystem. Besides, in this study we do not consider less popular projects since they have short development history and it is less probable to yield some patterns from these projects. However, the BC practices in less popular projects may still differ from those in popular projects. As for another threat, the Reid et al.'s dataset \cite{reid_2020_3898749} used in our study was released in 2020 (four years ago), and might not include popular JavaScript and TypeScript released after 2020. We manually checked 500 NPM libraries with the largest number of dependents and found that between them\footnote{We use the list in \url{https://leodog896.github.io/npm-rank/PACKAGES.html}, which is updated automatically (We accessed the page in September 2024).} only 16 were initially published after 2020. Hence Reid et al.’s dataset probably covers most popular JavaScript and TypeScript projects, and this threat might not affect the generalizability too much.

\textbf{Threats to conclusion validity} relate to insufficient high-quality BC commits to support our findings. To mitigate the threats, we try our best to extract breaking changes from commit messages, issues, pull requests and \textit{changelogs} to construct our breaking change dataset. We also carefully inspect our collected breaking changes and retain those related to JavaScript source code changes and explicitly documented by developers, which can help increase the quality of our dataset.

\section{Related Work}\label{section:related_work}

\subsection{Research on NPM Ecosystem}

A number of research works have investigated various aspects of the NPM ecosystem, which can help us understand the challenges in NPM ecosystem and provide future research directions. Decan et al.~\cite{decan2018impact} studied the impact of vulnerabilities in the NPM ecosystem by analyzing how and when security vulnerabilities are reported and fixed, and to which extent they affect other packages in the NPM ecosystem in the presence of dependency constraints. 
Liu et al.~\cite{liu2022demystifying} employed a knowledge graph method to characterize vulnerability propagation and evolution in the NPM ecosystem.
Cogo et al.~\cite{cogo2019empirical} conducted a systematic investigation into NPM dependency downgrades. 
Mujahid et al.~\cite{mujahid2023characteristics} summarized the characteristics of highly-selected projects in NPM. Maeprasart et al.~\cite{maeprasart2023understanding} illustrated the role of external pull requests in NPM project development.
Abdalkareem et al. \cite{abdalkareem2017developers} examined the usage of trivial packages, while Chen et al. \cite{chen2021helping} further uncovered the motivations driving JavaScript developers to publish such packages despite their potential drawbacks.
Venturini et al. \cite{venturini2023depended} studied the impact of manifesting breaking changes in NPM, which is the most closely related work to ours. 
In contrast to their study, our research delves into multiple dimensions of breaking changes, including the affected program elements, the impact on client applications, and the underlying reasons behind these changes. Additionally, we consider the specific language features of JavaScript and TypeScript at a fine-grained level.

\subsection{JavaScript Static Analysis}
Static analysis is a straightforward way to determine many properties of JavaScript programs and can be used to detect potential bugs and breaking API changes. Some open source tools like ESLint \cite{ESLint} and JSHint \cite{JSHint} support rule-based static analysis for JavaScript projects. They can be used for better code quality, such as making code follow JavaScript programming idioms. Some works proposed static analysis approaches from many aspects. For example, Madsen et al. built event-based call graphs \cite{madsen2015static} to detect event-related bugs by enhancing the static analysis framework JASI \cite{kashyap2014jsai} and TAJS \cite{jensen2011modeling}. Other works \cite{7372043, sung2016static, jensen2011modeling} studied static analysis for JavaScript in the DOM environment. Furthermore, in addressing the limitations of static analysis approaches for JavaScript, Chakraborty et al. presented a technique to supplement missing edges in the JavaScript call graph \cite{chakraborty2022automatic}. Through their research, they discovered that dynamic property access is a primary factor contributing to low recall in previous static analysis frameworks, typically missing some function invocations. By applying their proposed technique, they were able to improve the recall rate. 
However, due to the complexity of language features, a static analysis approach cannot cover all language use cases. Also, these static analysis approaches focus on the detection and elimination of ill-formed code, and optimization of the testing process, rather than breaking changes among commits. By applying their proposed technique, they were able to improve the recall rate. 

\subsection{Breaking Change Detection and Analysis on Other Platforms}
Besides JavaScript, a lot of studies focus on detecting and analyzing breaking changes for other programming languages, especially Java and Python, and their techniques may be learned by breaking change detection for JavaScript.
Brito et al. presented APIDiff \cite{brito2018apidiff} that can detect syntax-related breaking changes in Maven projects, such as method removal and visibility loss by reusing the refactoring detection tool RefDiff \cite{silva2017refdiff}. 
Some open-source tools can also check Java syntactic breaking changes, such as Clirr \cite{Clirr} and RevAPI \cite{RevAPI}.  For Python language, Du et al. proposed AexPy \cite{du2022aexpy} that can detect similar types of breaking changes like module removal and addition of required parameters, which extends the existing tool PyCompat \cite{zhang2020python} and {Pidiff} \cite{pidiff}.
Regarding non-syntactic breaking changes, to the best of our knowledge, Zhang et al. proposed Sembid \cite{zhang2022has} to detect behavioral breaking changes by measuring the semantic difference of call graphs between old and new programs: if a code change's semantic difference is larger than a threshold and not identified as {benign change}, it is classified as a BC. However, their used semantic diffs reflect the structural changes of programs, while many semantic breaking changes do not need many of those changes, e.g., only changing one condition, and adding a small change to the output string. Additionally, as the authors have pointed out, Sembid cannot distinguish semantic breaking changes from non-breaking changes such as re-implementation of a method. Therefore, directly adapting Sembid to JavaScript projects is not suitable.

Additionally, with the help of breaking change detection techniques, many works analyzed the impacts of breaking changes to downstream projects. For example, Jayasuriya et al. \cite{jayasuriya2023understanding, jayasuriya2024understanding} investigated the impacts of breaking changes (especially behavioral breaking changes) to client applications in the Maven ecosystem. They concluded with many findings, e.g., 11.58\% of the dependency updates contain breaking changes that impact the clients and 2.30\% version updates have behavioral breaking changes that impacted client tests. Similar to the findings in Venturini et al.'s study \cite{venturini2023depended}, while the breaking changes are only detected in a small number of code changes or version updates, they still cause negative impacts to downstream projects.

\section{Conclusion and Future Work}\label{section:conclusion}

In this study, for better comprehension of breaking changes in the NPM ecosystem and enhancing breaking change-related tools, we conducted an empirical study to bridge the knowledge gap from three aspects, with our carefully constructed breaking change dataset (1,519 breaking changes in total) sampled from a large set of popular NPM projects. We found that 95.4\% of the breaking changes detected by regression testing can be covered by developers' documentation, which proves that extracting breaking changes from documentation is reasonable. We then summarized the breaking changes in the NPM projects that are specific to JavaScript and TypeScript, and how developers make behavioral breaking changes, which yield many findings. Besides, we also investigated the reasons behind breaking changes in JavaScript and conclude with a taxonomy, which extend the previous works on motivations behind breaking changes. Based on our empirical findings, we provided actionable implications for future research, e.g., applying automatic renaming and naming techniques in JavaScript projects, and detecting code with similar functionalities, etc. 

In the future, we want to collect additional sources such as online discussions, and employ alternative approaches like dynamic analysis to gain a deeper understanding of breaking changes in the NPM ecosystem. We plan to investigate how to automatically identify the breaking changes in the NPM projects. We consider combining static analysis and dynamic analysis techniques to enhance the existing breaking change detection approaches. We also consider improving breaking change detection with the help of large language models since they show great performance improvements in many software engineering related tasks and can utilize the semantics in source code.

\begin{acks}
This research/project is supported by the National Science Foundation of China (No.62372398, No.72342025, and U20A20173), the Fundamental Research Funds for the Central Universities (No. 226-2022-00064), and the National Research Foundation, under its Investigatorship Grant (NRF-NRFI08-2022-0002). Any opinions, findings and conclusions or recommendations expressed in this material are those of the author(s) and do not reflect the views of National Research Foundation, Singapore.
\end{acks}

\bibliographystyle{ACM-Reference-Format}
\bibliography{main}


\end{document}